\documentclass[10pt,conference]{IEEEtran}
\usepackage{epsfig,rotating,setspace,latexsym,amsmath,epsf,amssymb,amsfonts,bm,theorem,cite,authblk, bbm,color, hyperref, multirow, caption, subcaption}
\IEEEoverridecommandlockouts
\allowdisplaybreaks

\title{Will 6G be Semantic Communications? Opportunities and Challenges from Task Oriented and Secure Communications to Integrated Sensing}
\begin{document}
\author[1]{Yalin E. Sagduyu}
\author[1]{Tugba Erpek}
\author[2]{Aylin Yener}
\author[3]{Sennur Ulukus}

\affil[1]{\normalsize  Virginia Tech, Arlington, VA, USA}

\affil[2]{\normalsize  The Ohio State University, Columbus, OH, USA}

\affil[3]{\normalsize University of Maryland, College Park, MD, USA}
\maketitle

\begin{abstract}
This paper explores opportunities and challenges of task (goal)-oriented and semantic communications for next-generation (NextG) communication networks through the integration of multi-task learning. This approach employs deep neural networks representing a dedicated encoder at the transmitter and multiple task-specific decoders at the receiver, collectively trained to handle diverse tasks including semantic information preservation, source input reconstruction, and integrated sensing and communications. To extend the applicability from point-to-point links to multi-receiver settings, we envision the deployment of decoders at various receivers, where decentralized learning addresses the challenges of communication load and privacy concerns, leveraging federated learning techniques that distribute model updates across decentralized nodes. However, the efficacy of this approach is contingent on the robustness of the employed deep learning models. We scrutinize potential vulnerabilities stemming from adversarial attacks during both training and testing phases. These attacks aim to manipulate both the inputs at the encoder at the transmitter and the signals received over the air on the receiver side, highlighting the importance of fortifying semantic communications against potential multi-domain exploits. Overall, the joint and robust design of task-oriented communications, semantic communications, and integrated sensing and communications in a multi-task learning framework emerges as the key enabler for context-aware, resource-efficient, and secure communications ultimately needed in NextG network systems.
\end{abstract}

\section{Introduction}
The landscape of wireless communications is undergoing a profound transformation, driven by the promise of next-generation (NextG) communications technologies and their potential applications. This transformation is strengthened by the ability to convey \emph{semantic information} as a fundamental cornerstone of NextG communication systems, introducing a paradigm shift that transcends conventional data transfer by embedding network communications with contextual depth and intelligence \cite{kountouris2021semantics, uysal2021semantic, gunduz2022beyond}. 

\emph{Task (goal)-oriented communications}, characterized by the utilization of deep neural networks (DNNs) at both the transmitter and receiver, aims to convey semantic information that is pertinent to the significance of a task, catering to the dynamic information needs of emerging applications powered by NextG communications \cite{shi2023task, sagduyu2023task}. Task-oriented communications involves training an encoder at the transmitter for source coding, channel coding, and modulation operations jointly with a decoder at the receiver to accomplish a specific task, such as classification (using the data samples that are originally available at the transmitter) without the need to transfer and reconstruct all data samples to the receiver \cite{xie2021deep,sagduyu2023task}. However, as we venture into the realm of 6G networks, the imperative extends beyond task completion to the preservation of semantic information along with information transfer, setting the stage for more intelligent, context-aware applications. 
\begin{figure*}[h]
	\centering	\includegraphics[width=1.7\columnwidth]{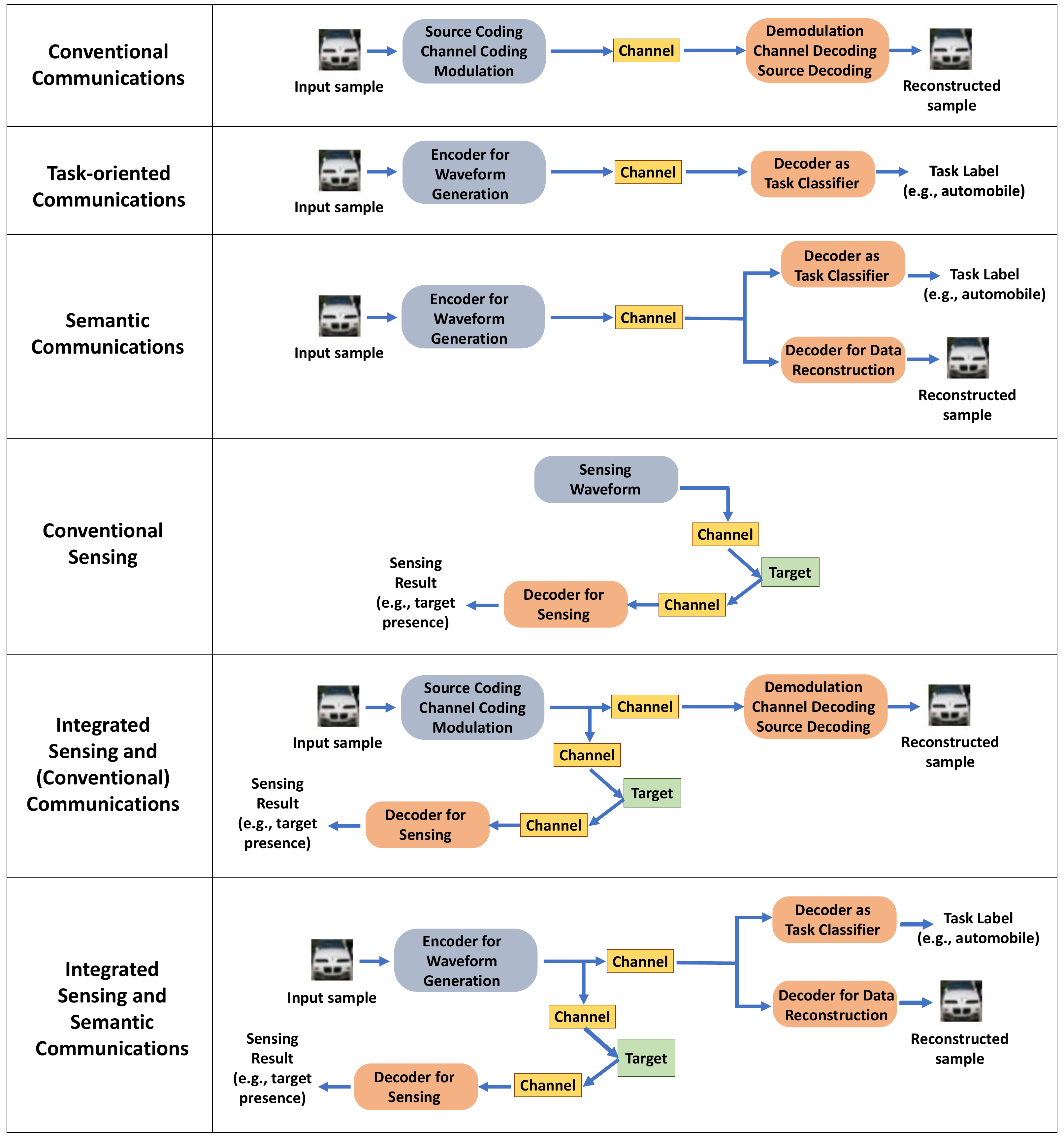}
	\caption{From conventional communications and sensing to task-oriented communications, semantic communications, and integrated sensing and communications.}
	\label{fig:IEEENetworkCases}
 \vspace{0.2cm}
\end{figure*}

\emph{Multi-task learning}, incorporating both reconstruction and semantic losses as part of the training process, stands as a catalyst for the design of \emph{semantic communications}. By combining the two objectives of reliable information reconstruction and the preservation of semantic meaning (captured by successful completion of a task at the receiver), multi-task learning lays the groundwork for semantic communication systems that transcend conventional data transfer \cite{sagduyu2022semantic}. The building blocks for transition from conventional communications to semantic communications is illustrated in Fig.\ref{fig:IEEENetworkCases}. This paradigm shift is poised to shape the trajectory of emerging 6G applications, such as augmented reality/virtual reality (AR/VR), Internet of Things (IoT), smart cities, vehicle-to-everything (V2X) networks, and autonomous driving. 
\begin{itemize}
    \item In AR/VR applications, semantic communications offers a gateway to more immersive and comprehensive experiences, providing a seamless entry into a realm where users not only observe but actively engage with digital environments. The preservation of semantic information such as patient attributes in telemedicine applications and failure points of machinery in remote maintenance applications ensures that the conveyed information in the AR/VR ecosystem retains its contextual significance, enhancing the quality and depth of AR/VR interactions. 
    \item For IoT and smart city applications, where interconnected sensors and actuators demand exchange of contextual information such as target types in surveillance feeds or operational states of machinery, semantic communications becomes instrumental in maintaining the integrity of information across diverse applications, from smart homes to industrial IoT. 
    \item In V2X network and autonomous driving applications, semantic communications contributes to the safety and efficiency of vehicular communication. The collaborative validation of semantic information across different tasks (potentially by multiple vehicles) such as classifying traffic signs or road conditions ensures a more holistic understanding of the environment, facilitating cooperative decision-making among vehicles. 
\end{itemize}

Multi-task learning of semantic communications can be further enriched by \emph{integrated (joint) sensing and communications} capability. As illustrated in Fig.\ref{fig:IEEENetworkCases}, transmitted signals are leveraged for dual purpose of sensing the environment \cite{liu2022integrated}. For that purpose, a decoder can detect and classify signals reflected from potential targets. To that end, a sensing loss can be combined with reconstruction and semantic losses for integrated sensing and semantic communications, ensuring resource efficiency by eliminating the need for separate probe signals for sensing the environment  \cite{sagduyu2023joint}. 
\begin{figure*}[h!]
	\centering	\includegraphics[width=1.8\columnwidth]{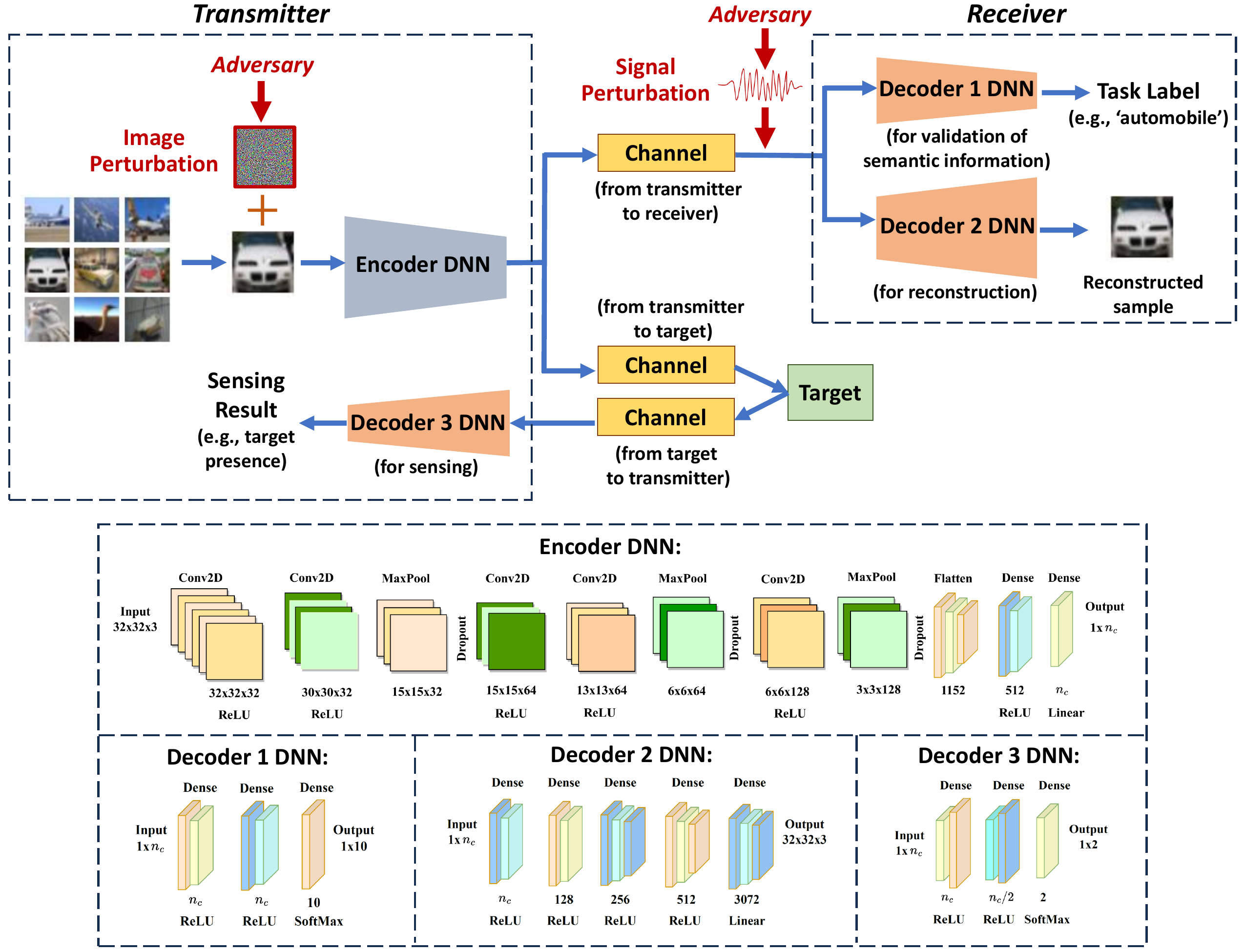}
	\caption{System model of multi-task learning for task-oriented communications, semantic communications, and integrated sensing and communications.}
	\label{fig:IEEENetworkSystem}
 \vspace{0.2cm}
\end{figure*}

Extending beyond traditional point-to-point links, there is a need to adapt semantic communications for \emph{multi-receiver communications} with different tasks to be completed at different receivers \cite{sagduyu2023multi}. By deploying decoders at various receivers for different tasks, this extension needs to embrace \emph{decentralized learning} to address challenges associated with communication load and privacy concerns. Leveraging \emph{federated learning}, the multi-task learning process can be distributed across decentralized nodes, enabling a collaborative learning process \cite{mortaheb2022personalized}.

Yet, amidst these advancements, the \emph{security} challenges associated with semantic communications cannot be overlooked. DNNs are known to be highly vulnerable to even slight variations in training and test data samples \cite{Adesina2022}. These variations can be deliberately added as small perturbations by adversaries, leading to \emph{adversarial machine learning} threats that aim to disrupt the training and test processes of the underlying DNNs. To that end, stealth but powerful attacks can be launched effectively to manipulate both inputs at the encoder and signals received over the air, underscoring the critical importance of security measures to safeguard semantic communications against these \emph{multi-domain attacks} \cite{sagduyu2023task, sagduyu2022semantic, sagduyu2022vulnerabilities}.

As we explore the realms of task-oriented communications, semantic communications, and integrated sensing and communications, this paper navigates the intricate landscape of NextG networks, providing insights and strategies to propel the context-awareness, reliability, resource efficiency, and security of communication systems in the era of 6G. We demonstrate the effectiveness of pursuing multiple objectives of preserving the semantic information, reconstructing the data samples, and performing the environment sensing during the transfer of CIFAR-10 image samples over additive white Gaussian noise (AWGN) and Rayleigh channels, and highlight the potential vulnerabilities to stealth attacks built upon adversarial machine learning.

The remainder of the paper is organized as follows. Secs.~\ref{sec:TOC}, \ref{sec:semantic}, and \ref{sec:ISC} present task-oriented communications, semantic communications, and integrated sensing and communications, respectively, in a multi-task learning framework. Sec.~\ref{sec:FL} describes extension to distributed learning. Sec.~\ref{sec:AML} presents vulnerabilities of semantic communications to adversarial machine learning threats. Sec.~\ref{sec:conclusion} concludes the paper by providing a summary of contributions and describing future research directions. 

\section{Task-Oriented Communications} \label{sec:TOC}
Information systems rely on reliable and efficient processing of information encompassing generation, feature extraction, compression, and transfer of different data modalities. Traditionally, wireless systems are built on the principle of ensuring reliable communications subject to channel impairments. Consequently, they often overlook the significance of transferred information in relation to the underlying tasks or goals. 

To unlock the full potential of emerging applications promised by 6G, the paradigm of \emph{task-oriented communications} offers a transformative approach. Here, the primary objective is not necessarily reconstructing data samples but successfully completing specific tasks (e.g., classifying the received signals to labels of interest) at a receiver by utilizing the available data at the transmitter. One of the primary imperatives for 6G networks is to cater to a diverse range of applications, from AR/VR and IoT to V2X and autonomous driving. Task-oriented communications is uniquely poised to meet the distinct demands of these applications by prioritizing the successful execution of specific tasks over the meticulous reconstruction of transmitted information.

The overall system model is shown in Fig.~\ref{fig:IEEENetworkSystem}. In task-oriented communications, the transmitter's operations, encompassing source coding, channel coding, and modulation operations, are conceptualized as an encoder, namely a DNN, tasked with generating and transmitting low-dimensional feature vectors. Unlike the conventional receiver chain, which focuses on information reconstruction, the task-oriented receiver employs a dedicated decoder to directly perform the assigned task, such as classifying the received signals, while obviating the need for the resource-intensive process of reconstructing the original input samples. 

This shift in perspective not only redefines the fundamental architecture of communication systems by distributing intelligence between the encoder and decoder, and streamlining the communication 
process but also holds the promise of remarkable efficiency gains pertinent to the demanding requirements of 6G applications. Task-oriented communications reduces the number of transmissions, latency, and energy consumption by limiting the size of encoder output. These gains are especially critical for emerging 6G applications requiring ultra-low latency for split-second decisions such as needed in AR/VR and V2X systems, and high energy efficiency for battery-powered devices such as deployed in IoT systems.

We demonstrate the anticipated benefits of task-oriented communications for the transfer of necessary information of reduced dimension to a receiver to complete an image classification task. For that purpose, we consider images from CIFAR-10 dataset available at the transmitter. This dataset consists of 60,000 color images, each of size 32$\times$32$\times$3, with a split of 50,000 and 10,000 samples for training and testing. The transmitter encodes these images and transmits their latent representations of reduced dimension over a wireless channel. This dimension is controlled by the encoder's output size, $n_c$, which is an important design parameter that determines the number of channel uses and the input size of decoders. For example, when $n_c$ is 20, the dimension of transferred information is reduced from 32$\times$32$\times$3 = 3072 to 20 with the compression ratio of 0.65\%. We consider two channel models, namely AWGN and Rayleigh channels. The decoder at the receiver classifies the received signals to one of the 10 labels. 

The synergistic relationship between the encoder and the decoder is modeled as an end-to-end DNN that is trained by considering channel and data characteristics to optimize task performance. The success of this optimization is quantified via machine learning metrics, such as the loss function in a classification task. To that end, the encoder-decoder pair is jointly trained with the categorical cross-entropy as the loss function. 

The DNN architectures are determined by gradually increasing the number of layers and layer sizes until the average classification accuracy cannot be further improved as the performance measure of task completion. As shown in Fig.~\ref{fig:IEEENetworkSystem}, layers of the encoder are Input (size = 32$\times$32$\times$3), Conv2D (filter size = 32), MaxPooling2D, Dropout, Conv2D (filter size = 64), MaxPooling2D, Dropout, Conv2D (filter size = 128),  MaxPooling2D, Dropout, Flatten, Dense (size = 512), Dropout, and Dense (size = $n_c$). Conv2D's kernel size is (3, 3), MaxPooling2D’s pool size is (2,2), dropout rate is 0.25, and ReLU is the activation function of hidden layers. The size of encoder's output layer is $n_c$ and its activation function is Linear. Layers of the decoder for semantic information validation are two Dense layers of size $n_c$, followed up by a Dense layer of size 10 as the output layer, where the activation function is Softmax. 

\begin{figure}[h!]
	\centering
	\includegraphics[width=\columnwidth]{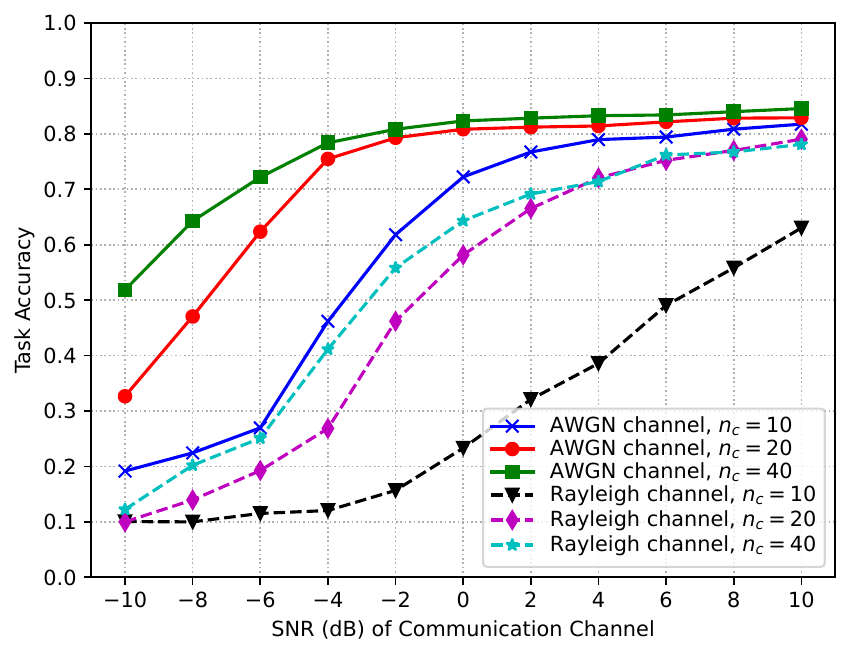}
	\caption{Task accuracy vs. SNR (dB) of communication channel.}
	\label{fig:NetworkMagAccuracySNR}
 \vspace{0.2cm}
\end{figure}

Fig.~\ref{fig:NetworkMagAccuracySNR} shows how the task accuracy improves when the signal-to-noise ratio (SNR) of communication channel or $n_C$ increases. While the task accuracy is lower under AWGN channel compared to Rayleigh channel in general, the performance gap closes with higher SNR or $n_c$. 

\section{Semantic Communications} \label{sec:semantic}
In traditional communication systems, the central objective revolves around the reliable transmission of messages, accounting for channel impairments to minimize a reconstruction loss, such as the symbol error rate. This involves the design of individual or collaborative operations at the transmitter and receiver. DNNs play a pivotal role in this framework, acting as \emph{autoencoders} that comprehensively capture transmitter and receiver operations, including source coding, channel coding and modulation at the transmitter, and demodulation, channel decoding and source decoding at the receiver. The overarching aim is to minimize the end-to-end reconstruction loss.

In addition to the reliable reconstruction of information, \emph{semantic communications} emerges as a paradigm shift, striving to preserve the intrinsic meaning of the conveyed information throughout the communication process to the receiver. In semantic communications, the training loss for the autoencoder encompasses not only the reconstruction loss inherent in traditional communications but also introduces the concept of \emph{semantic loss}. The semantic loss aims to capture the preservation of meaning during information transfer, ensuring that the conveyed information retains its intended significance. For that purpose, the semantic loss is measured by the performance of completing a mission-critical task  with the conveyed information. 

As shown in Fig.~\ref{fig:IEEENetworkSystem}, an encoder is employed at the transmitter to handle source coding, channel coding, and modulation, whereas the setup at the receiver involves two decoders. One decoder focuses on reconstructing information through a process involving joint demodulation, channel decoding, and source decoding. Simultaneously, another decoder evaluates whether semantic information is preserved, effectively performing a machine learning task such as image classification. The reconstruction loss is combined with a semantic loss by jointly training three DNNs, namely one encoder and two decoders. This method adeptly captures latent feature spaces, facilitating the reliable transfer of compressed feature vectors with minimal channel usage, all while maintaining a low semantic loss.

The exploration of semantic communications spans diverse data types, ranging from text, image, and video to speech and audio. The goal is to maintain the integrity of meaning across different forms of information, aligning seamlessly with the applications of 6G networks that will benefit significantly from the preservation of semantic meaning, enriching the user experience and enabling more contextually aware communication.

Building upon the demonstration use case from Sec.~\ref{sec:TOC}, we add mean-squared error as the reconstruction loss (to recover image samples at the receiver) in addition to the semantic loss (to validate if the semantic information is conveyed) in a multi-task learning framework, where we can set individual weights for losses. As shown in Fig.~\ref{fig:IEEENetworkSystem}, layers of the decoder for information reconstruction are  Dense layers of size $n_c$, 128, 256, 512, and 3072 (activation function is ReLU for the first three layers and Linear for the last layer), followed by a Reshape layer that maps the dimension from 3072 to 32$\times$32$\times$3. 

\begin{figure}[h]
	\centering
	\includegraphics[width=\columnwidth]{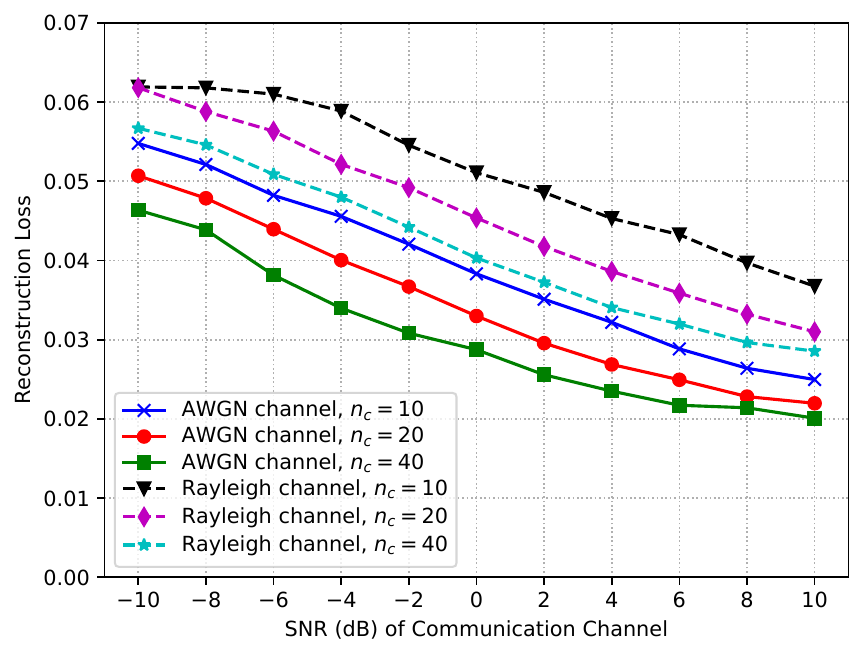}
	\caption{Reconstruction loss vs. SNR (dB) of communication channel.}
	\label{fig:NetworkMagReconsSNR}
 \vspace{0.2cm}
\end{figure}

Fig.~\ref{fig:NetworkMagReconsSNR} shows how the reconstruction loss decreases when increase the SNR or $n_c$ under both AWGN and Rayleigh channel models, leading to more reliable recovery of image samples. On the other hand, multi-task learning can help the task accuracy for semantic information validation reach the performance levels presented in Sec.~\ref{sec:TOC}.

\section{Integrated Sensing and Communications} \label{sec:ISC}
The integration of sensing functionalities emerges as a pivotal component in NextG communications systems, expanding beyond traditional data transmissions to include the \emph{integrated sensing and communications} paradigm. This section explores the formulation of sensing as a task in multi-task learning for semantic communications, where the extended set of tasks includes not only   preservation of semantic information and reconstruction of input samples but also environment sensing, namely detection of a potential target.

Integrated sensing and communications leverages the dual use of communication signals for sensing and eliminates the need for additional probe signals, thereby conserving bandwidth, reducing delay, and enhancing energy efficiency. In the context of 6G applications, this resource-efficient integration of sensing into communication capabilities unveils transformative possibilities. One key example is in the realm of V2X networks. The dual functionality of communication and sensing allows vehicles to exchange crucial information about their positions and intentions, while also continuously sensing and monitoring the surroundings for potential hazards and dynamically changing conditions on the road, and detecting objects such as other vehicles, pedestrians, cyclists, traffic signs, and obstacles like road barriers.

Integrated sensing and communications presents two distinctive avenues: one involves repurposing existing communication waveforms for sensing, as seen in WiFi sensing. On the other hand, a more forward-thinking alternative is to design a communication waveform with a dual purpose. In this approach, the waveform is crafted not solely for information exchange but also with a specific focus on sensing, especially in tasks like target detection. This strategy ensures that the designed waveform is tailored to excel in both communication and sensing tasks, maximizing their effectiveness in diverse scenarios.

To optimize the overall performance, we consider a DNN-driven design of communication waveform shown in Fig.~\ref{fig:IEEENetworkSystem} for both information transfer and sensing. Employing an encoder, the transmitter conducts joint operations encompassing source coding, channel coding, and modulation to generate signals. On the other end, the receiver employs another DNN, serving as a decoder, for joint operations including demodulation, channel decoding, and source decoding to reconstruct data samples. In the meantime, the transmitted signal has a dual role, facilitating communication with the receiver and enabling sensing capabilities. In the presence of a target, the signal is reflected from the target and potentially received by the transmitter or a receiver, leading to two possible cases of sensing. In \emph{monostatic sensing}, the transmitter performs the sensing operation based on the signals reflected from the target. In \emph{bistatic sensing}, the transmitter of signals and the receiving end of reflected signals are separated such that the receiving end could be the receiver of communication signals or any other receiver deployed for environment sensing. 

In all these cases, an additional  decoder is employed either at the transmitter or at a receiver for sensing purposes. This decoder can be tasked with detecting the target's presence and determining its range. Through multi-task learning, all involved DNNs undergo joint training, considering data, channel, and potential target characteristics. In this multi-task learning framework, semantic communications can be further incorporated by using an additional decoder at the receiver, operating as a task classifier. This decoder assesses the fidelity of task completion such as image classification based on received signals, enhancing the incorporation of semantics into the joint sensing and communication process.

\begin{figure}[h]
	\centering
	\includegraphics[width=\columnwidth]{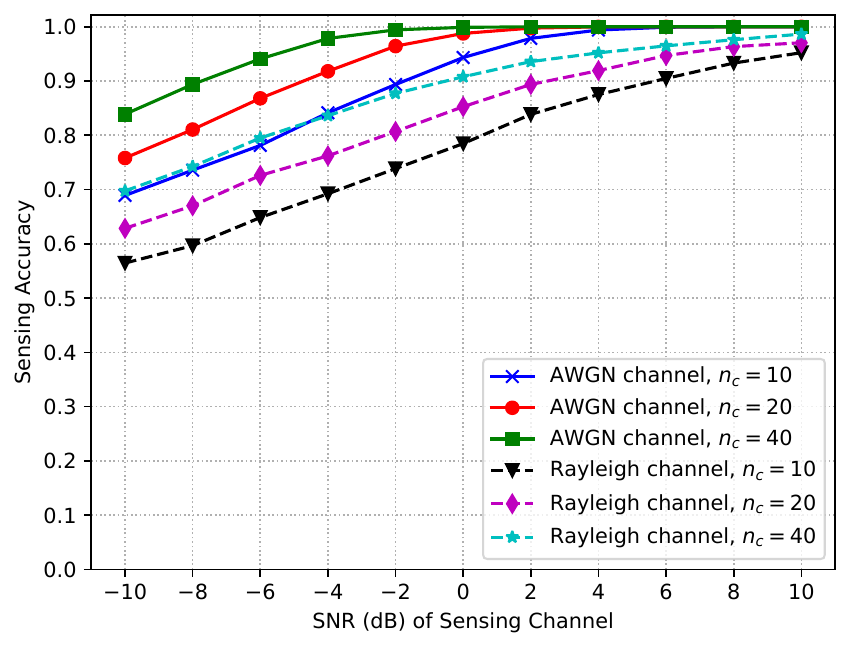}
	\caption{Sensing accuracy vs. SNR (dB) of sensing channel (corresponding to channels to and from the target).}
	\label{fig:NetworkMagSensingAccuracySNR}
 \vspace{0.2cm}
\end{figure}

Building upon the multi-task learning framework from Sec.~\ref{sec:semantic}, we measure the sensing loss as categorical cross-entropy in addition to reconstruction and semantic losses (discussed in Secs.~\ref{sec:TOC} and \ref{sec:semantic}) such that all the underlying DNNs are jointly trained according to the weights assigned to the underlying losses in multi-task learning. As shown in Fig.~\ref{fig:IEEENetworkSystem}, layers of the decoder for sensing are Dense layers of size $n_c$, $n_c/2$, and 2, where the activation function is ReLU for the first two layers and Softmax for the output layer. Considering the mono-static sensing case, Fig.~\ref{fig:NetworkMagSensingAccuracySNR} shows how the sensing accuracy (namely, the probability of successfully detecting the presence or absence of a target) increases with the SNR of channels to and from the target, whereas the task accuracy and reconstruction loss can reach the performance levels presented in Secs.~\ref{sec:TOC} and \ref{sec:semantic}, respectively, through multi-task learning. 

\section{Multi-user Communications and Distributed Learning} \label{sec:FL}
The extension of semantic communication to multiple users introduces a paradigm shift, transforming the traditional point-to-point communications into a collaborative model via a distributed framework. This multi-user semantic communications approach not only accommodates the diverse information needs of different users but also facilitates the validation of semantic information by multiple receivers, each potentially tasked with distinct objectives.

\noindent \textbf{Extension of semantic communications to multiple receivers.} The studies of semantic communications have initially focused on preserving the meaning of information for a single receiver. In the multi-user context, this approach can be expanded to cater to diverse recipients, each assigned a unique task. Each receiver, equipped with a specific decoder tailored to its task, validates the semantic information conveyed. This extension allows for the simultaneous fulfillment of multiple tasks, ranging from classification to contextual understanding, leveraging semantic information to its fullest potential.

The need for \emph{multi-user semantic communications} arises from the diverse information requirements inherent in advanced applications of 6G. In collaborative decision-making, various users possess distinct semantic tasks crucial for comprehensive information understanding. 6G's vision encompasses a multitude of applications, such as AR/VR, IoT, smart city, V2X networks and autonomous driving, requiring a communication system that can cater to varied and nuanced semantic information needs. Multi-user semantic communication becomes essential to ensure that each participant receives information tailored to their specific tasks, fostering a more intelligent and responsive 6G ecosystem.

\noindent \textbf{Multi-task learning in a collaborative setting.}
Ensuring that each receiver's semantic task is adequately addressed while preserving global semantic coherence poses a significant technical challenge. To achieve multi-user semantic communications, the concept of multi-task learning can be extended to a \emph{distributed and collaborative learning} setting. Each receiver's task becomes a distinct facet of the overarching communication objective. The encoder at the transmitter is trained to accommodate these diverse tasks, and each decoder at the receiver is tailored to validate the semantic information within the context of its designated task.
 
\noindent \textbf{Transition to federated learning.} 
To address the distributed nature of multi-user semantic communication, the evolution from multi-task learning to \emph{federated learning} becomes pivotal. Federated learning enables decentralized training across multiple receivers, optimizing the learning process by leveraging local datasets at each receiver while collectively enhancing semantic information across the entire network.

Federated learning provides a powerful set of advantages for multi-user semantic communications in 6G applications. Its privacy-preserving nature ensures that sensitive information remains on user devices, addressing data sharing concerns. Enabling decentralized model training optimizes resource utilization, seamlessly integrating with edge computing for enhanced responsiveness in applications like AR/VR, IoT, smart city, V2X networks, and autonomous driving. Federated learning's adaptability to diverse data distributions ensures robust and scalable models, crucial in multi-user semantic communication. Additionally, the reduction in exchanged network data enhances efficiency, a vital aspect in 6G scenarios.

\section{Security Vulnerabilities} \label{sec:AML}
With the reliance on DNNs, semantic communications becomes susceptible to \emph{adversarial machine learning threats}. This section explores adversarial, poisoning, and backdoor attacks in two critical scenarios: manipulation of the input to the transmitter's encoder and manipulation of the over-the-air signal received by the decoder at the receiver.

\noindent \textbf{Adversarial attacks.}
Adversarial attacks target the robustness of the DNNs by strategically tampering with test data. Adversaries may manipulate the input to the encoder at the transmitter or interfere with signals received by the decoder over the air. In both cases, subtle perturbations can be introduced to deceive the learning process, leading to the distortion of semantic information, information reconstruction, and sensing outcome. We can distinguish targeted and untargeted adversarial attacks. In a targeted adversarial attack, adversaries manipulate the model's output with the goal of causing misclassification into a specific, predefined target class. Conversely, in an untargeted adversarial attack, adversaries introduce perturbations into the input data in a way that causes any misclassification without steering the decision towards a predefined target.

\noindent \textbf{Poisoning attacks.}
Poisoning attacks focus on compromising the integrity of DNNs by injecting malicious data into training process. Adversaries may tamper with the training dataset of the transmitter at the input of the encoder or  may strategically interfere with and manipulate over-the-air signals that are collected for training. These attacks aim to distort the learning representations, resulting in biased semantic features and leading to incorrect task executions at multiple decoders.

\noindent \textbf{Backdoor attacks.}
Backdoor (or Trojan) attacks insert hidden triggers to the training data such as manipulating some training data samples and falsifying their labels. Adversaries may embed subtle backdoors by manipulating some inputs to the encoder during training. Adversaries may also insert triggers such as phase shifts to some wireless signals transmitted over a channel. These triggers are later activated in test time selectively in some instances. This stealth attack can result in manipulated task outcomes at different decoders for task classification, information reconstruction, and sensing purposes.

\noindent \textbf{Mitigation strategies.}
In response to the multifaceted security threats, defense strategies are needed to protect task-oriented and semantic communications, and their extension with joint sensing.

\begin{itemize}
    \item Proactive defenses, including adversarial training, randomized smoothing, and certified methods, enhance model resilience against adversarial attacks. Continual model audits and anomaly detection mechanisms contribute to identifying and mitigating adversarial manipulations.
    
    \item Defense against poisoning attacks involves data validation, robust outlier detection, and leveraging decentralized learning like federated learning to reduce the impact of poisoned data. Continuous monitoring of semantic representations ensures the integrity of learned features.
    
    \item Backdoor threat mitigation entails measures such as input validation, adversarial training, and anomaly detection for over-the-air signal manipulations. Regular updates and diversification of the training dataset reduce the risk of backdoor insertion.
\end{itemize}

To demonstrate security vulnerabilities, we consider adversarial attacks that add small perturbations to the encoder inputs to cause potential errors in the classification task that checks whether the semantic information is preserved.  
We apply the Fast Gradient Sign Method (FGSM) to generate perturbations for a white-box and untargeted adversarial attack. By calculating the gradient of the model's loss with respect to the input, the adversary generates adversarial examples by perturbing the input data in the direction that maximizes the semantic loss, aiming to mislead the model's predictions for semantic information validation. For a stealthy attack, we suppose that the strength of perturbation added to input samples is limited by a given perturbation-to-signal ratio (PSR). 

\begin{figure}[h]
	\centering
	\includegraphics[width=1\columnwidth]{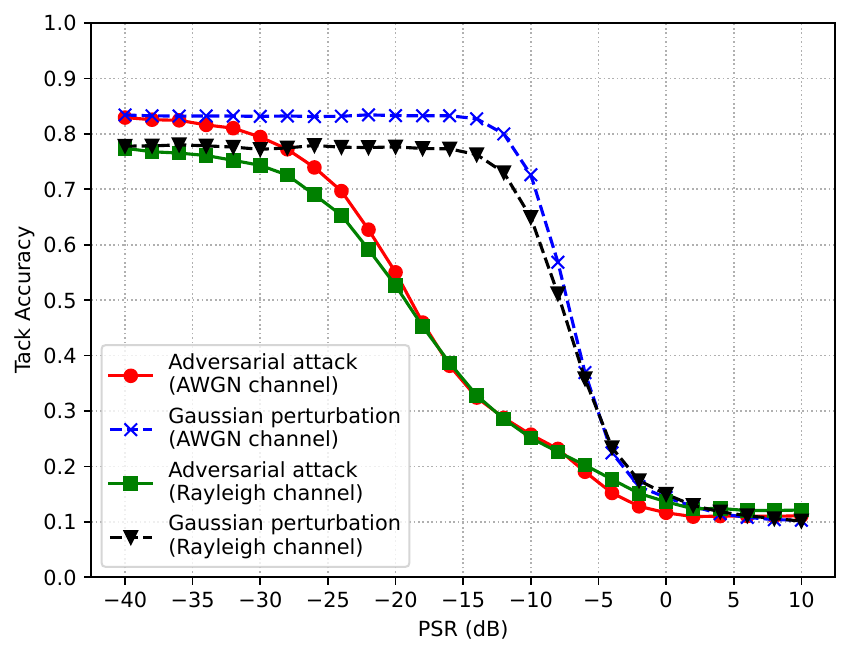}
	\caption{Task accuracy vs. PSR (SNR = 10dB, $n_c$ = 20) under adversarial attack.}
	\label{fig:NetworkMagSecurityaccuracy}
 \vspace{0.2cm}
\end{figure}

As shown in Fig.~\ref{fig:NetworkMagSecurityaccuracy}, as the PSR increases, the task accuracy quickly drops leading to excessive number of classification errors. On the other hand, the use of Gaussian noise as perturbation has lower attack performance compared to adversarial attack, and its effect remains limited unless the PSR is very high. This attack example is built upon a white-box attack, where the adversary possesses complete knowledge of the target model. In contrast, a black-box attack can be launched by involving no information about the model's internal structure. Gray-box attacks fall in between, with partial knowledge. These variations can be studied to assess the attack surface exploited by different types of adversaries, each possessing different levels of knowledge.

\section{Conclusion and Future Research Directions} \label{sec:conclusion}
The convergence of semantic communications with task orientation and integrated sensing in multi-task learning holds the key to unlocking the transformative potential of context-aware, scalable, and secure communications in the era of 6G networks. This synergistic alignment not only optimizes information transfer efficiency but also establishes a foundation for intelligent, context-aware communications. Task-oriented communications prioritizes successful task completion over conventional information reconstruction, crucial for addressing the diverse demands of 6G applications. Semantic communications contributes an additional layer of depth of reliable data reconstruction in addition to preserving semantic information, catering to the requirements of emerging 6G applications. Sensing can be added as another task to this multi-task learning framework by efficiently utilizing transmitted signals for both communication and sensing purposes. Distributed operation with multiple receivers via federated learning ensures collaborative model training, addressing communication load and privacy concerns for a more adaptive and scalable system.  However, security implications due to adversarial machine learning underscore the imperative to fortify task-oriented communications, semantic communications, and integrated sensing and communications against potential exploits.

\emph{Future research directions} include incorporation of \emph{timeliness} objectives along with preserving semantic information. The information freshness aspect, introducing metrics like the \emph{age of information} (AoI) \cite{yates2021age}, is especially critical for 6G applications with fast decision-making mandates. To that end, novel metrics like \emph{age of task information} and \emph{age of semantic information} are needed for completing tasks and conveying semantic information in a timely manner \cite{sagduyu2023age}.  The imperative to enhance \emph{privacy} in the presented multi-task learning framework suggests several avenues for future research to prevent eavesdroppers from learning outcomes of different tasks and capturing semantic information. Cryptographic techniques such as homomorphic encryption, secure multiparty computation, and differential privacy techniques can protect multi-task learning against eavesdropping. Additionally, tailoring federated learning protocols for multi-task scenarios and exploring adversarial machine learning techniques to fool the deep learning-empowered eavesdroppers offer promising approaches to fortify privacy measures. Investigation into these research directions will contribute to maximizing the transformative potential of multi-task learning for semantic communications along with task-orientation and joint sensing in NextG systems.

\bibliographystyle{IEEEtran}
\bibliography{references}

\begin{thebibliography}{10}
\providecommand{\url}[1]{#1}
\csname url@samestyle\endcsname
\providecommand{\newblock}{\relax}
\providecommand{\bibinfo}[2]{#2}
\providecommand{\BIBentrySTDinterwordspacing}{\spaceskip=0pt\relax}
\providecommand{\BIBentryALTinterwordstretchfactor}{4}
\providecommand{\BIBentryALTinterwordspacing}{\spaceskip=\fontdimen2\font plus
\BIBentryALTinterwordstretchfactor\fontdimen3\font minus \fontdimen4\font\relax}
\providecommand{\BIBforeignlanguage}[2]{{%
\expandafter\ifx\csname l@#1\endcsname\relax
\typeout{** WARNING: IEEEtran.bst: No hyphenation pattern has been}%
\typeout{** loaded for the language `#1'. Using the pattern for}%
\typeout{** the default language instead.}%
\else
\language=\csname l@#1\endcsname
\fi
#2}}
\providecommand{\BIBdecl}{\relax}
\BIBdecl

\bibitem{kountouris2021semantics}
M.~Kountouris and N.~Pappas, ``Semantics-empowered communication for networked intelligent systems,'' \emph{IEEE Communications Magazine}, vol.~59, no.~6, pp. 96--102, 2021.

\bibitem{uysal2021semantic}
E.~Uysal, O.~Kaya, A.~Ephremides, J.~Gross, M.~Codreanu, P.~Popovski, M.~Assaad, G.~Liva, A.~Munari, T.~Soleymani, B.~S. Soret, and H.~Johansson, ``Semantic communications in networked systems,'' \emph{IEEE Network}, vol.~36, no.~4, pp. 233--240, 2022.

\bibitem{gunduz2022beyond}
D.~G{\"u}nd{\"u}z, Z.~Qin, I.~E. Aguerri, H.~S. Dhillon, Z.~Yang, A.~Yener, K.~K. Wong, and C.-B. Chae, ``Beyond transmitting bits: Context, semantics, and task-oriented communications,'' \emph{IEEE Journal on Selected Areas in Communications}, vol.~41, no.~1, pp. 5--41, 2023.

\bibitem{shi2023task}
Y.~Shi, Y.~Zhou, D.~Wen, Y.~Wu, C.~Jiang, and K.~B. Letaief, ``{Task-Oriented Communications for 6G: Vision, Principles, and Technologies},'' \emph{IEEE Wireless Communications}, vol.~30, no.~3, pp. 78--85, 2023.

\bibitem{sagduyu2023task}
Y.~E. Sagduyu, S.~Ulukus, and A.~Yener, ``Task-oriented communications for {nextG}: End-to-end deep learning and {AI} security aspects,'' \emph{IEEE Wireless Communications}, vol.~30, no.~3, pp. 52--60, 2023.

\bibitem{xie2021deep}
H.~Xie, Z.~Qin, G.~Y. Li, and B.-H. Juang, ``Deep learning enabled semantic communication systems,'' \emph{IEEE Transactions on Signal Processing}, vol.~69, pp. 2663--2675, 2021.

\bibitem{sagduyu2022semantic}
Y.~E. Sagduyu, T.~Erpek, S.~Ulukus, and A.~Yener, ``Is semantic communication secure? a tale of multi-domain adversarial attacks,'' \emph{IEEE Communications Magazine}, vol.~61, no.~11, pp. 50--55, 2023.

\bibitem{liu2022integrated}
F.~Liu, Y.~Cui, C.~Masouros, J.~Xu, T.~X. Han, Y.~C. Eldar, and S.~Buzzi, ``Integrated sensing and communications: Toward dual-functional wireless networks for {6G} and beyond,'' \emph{IEEE Journal on Selected Areas in Communications}, vol.~40, no.~6, pp. 1728--1767, 2022.

\bibitem{sagduyu2023joint}
Y.~E. Sagduyu, T.~Erpek, A.~Yener, and S.~Ulukus, ``Joint sensing and semantic communications with multi-task deep learning,'' \emph{arXiv preprint arXiv:2311.05017}, 2023.

\bibitem{sagduyu2023multi}
------, ``Multi-receiver task-oriented communications via multi-task deep learning,'' in \emph{IEEE Future Networks World Forum}, 2023.

\bibitem{mortaheb2022personalized}
M.~Mortaheb, C.~Vahapoglu, and S.~Ulukus, ``Personalized federated multi-task learning over wireless fading channels,'' \emph{Algorithms}, vol.~15, no.~11, p. 421, 2022.

\bibitem{Adesina2022}
D.~Adesina, C.-C. Hsieh, Y.~E. Sagduyu, and L.~Qian, ``Adversarial machine learning in wireless communications using {RF} data: A review,'' \emph{IEEE Communications Surveys \& Tutorials}, vol.~25, no.~1, pp. 77--100, 2023.

\bibitem{sagduyu2022vulnerabilities}
Y.~E. Sagduyu, T.~Erpek, S.~Ulukus, and A.~Yener, ``Vulnerabilities of deep learning-driven semantic communications to backdoor (trojan) attacks,'' in \emph{Conference on Information Sciences and Systems (CISS)}, 2023.

\bibitem{yates2021age}
R.~D. Yates, Y.~Sun, D.~R. Brown, S.~K. Kaul, E.~Modiano, and S.~Ulukus, ``Age of information: An introduction and survey,'' \emph{IEEE Journal on Selected Areas in Communications}, vol.~39, no.~5, pp. 1183--1210, 2021.

\bibitem{sagduyu2023age}
Y.~E. Sagduyu, S.~Ulukus, and A.~Yener, ``Age of information in deep learning-driven task-oriented communications,'' in \emph{IEEE INFOCOM Age of Information Workshop}, 2023.

\end{thebibliography}

\end{document}